\documentclass[dvips]{article}

\usepackage{icrc2011}

\title{Search for Nuclei Sources in the Ultra-High Energy Cosmic Ray Data}

\newcommand{\etal}{\MakeLowercase{\textit{et al. }}} 
\shorttitle{G.~GIACINTI \etal Search for UHE Nuclei Sources}

\authors{G.~Giacinti$^{1}$, D.~V.~Semikoz$^{1,2}$}
\afiliations{$^1$AstroParticle and Cosmology (APC), Paris, France\\ $^2$Institute for Nuclear Research of the Russian Academy of Sciences, Moscow, Russia}
\email{gwenael.giacinti@apc.univ-paris7.fr}

\abstract{We present a new method to search for heavy nuclei sources, on top of background, in the Ultra-High Energy Cosmic Ray data~\cite{Giacinti:2010ep}. We apply it to the 69 events with energies $E \geq 55$\,EeV published by the Pierre Auger Collaboration. We find a set of events for which the method reconstructs the source near the Virgo galaxy cluster. The probability to have a comparable set of events in some background is $\sim 0.7$\%. The reconstructed source is located at $\simeq 8.5$ degrees from the active galaxy M87. The probability to reconstruct the source at less than 10 degrees from M87 for data already containing a comparable set of events is $\sim 0.4$\%. This may be a hint at the Virgo galaxy cluster as an ultra-high energy heavy nuclei source~\cite{Semikoz:2010cc,Giacinti:2010ep}. We discuss the capability of current and near future experiments to test this possibility. Such a scenario gives a self-consistent description of the Auger anisotropy and composition data at the highest energies.}
\keywords{Ultra-High Energy Cosmic Rays, Galactic Magnetic Field}

\begin{document}
\maketitle


\section{Introduction}

The Pierre Auger Observatory measurements of the Ultra-High Energy Cosmic Ray (UHECR) composition are compatible with a shift towards heavy nuclei, above $\sim 10^{19}$\,eV~\cite{Collaboration:2010yv}. The analysis of the muon data from the Yakutsk EAS Array also hint at a significant fraction of heavy nuclei at the highest energies~\cite{Glushkov:2007gd}. On the contrary, both HiRes measurements~\cite{BelzICRC} and preliminary results of Telescope Array~\cite{TA} are still consistent with a proton composition.

Both for protons and iron nuclei, UHECR sources must be located in the local Universe, within $r \leq 150$\,Mpc for energies $E>6 \times 10^{19}$\,eV. They should lie within the Large Scale Structure (LSS) of galaxy distribution.

Until now, methods to look for sources have been proposed for proton or light nuclei primaries~\cite{Golup:2009cv,Giacinti:2009fy}. We present here a new method to search for UHE heavy nuclei sources. In most regions of the sky, one cannot detect heavy nuclei sources without a better knowledge of the Galactic Magnetic Field (GMF) than currently available~\cite{Giacinti:2010dk,Giacinti:2011uj,ICRC_0229}. Nonetheless, we show that in some favourable cases, one may construct an algorithm to detect some of such sources with the present knowledge of the GMF.

We apply this method to the 69 Auger events published in Ref.~\cite{Collaboration:2010zzj}. We detect a set of events for which the associated reconstructed source lies near Virgo. The Virgo cluster is in principle a good candidate for containing one or several UHECR source(s). The probability to have a comparable set of events in some background and reconstruct the source in any direction of the sky is $\sim 7\times10^{-3}$. If such a set of events already exists in the data, the probability to reconstruct the source at less than 10$^{\circ}$ from M87 is $\sim 4\times10^{-3}$. The possibility that the detected feature is the heavy nuclei image of Virgo would be compatible with both the Auger composition and anisotropy at the highest energies. However, the present statistics do not allow us to conclude firmly. We discuss the ability of current and near future experiments to confirm or rule out this possibility.

We present the method in Section~\ref{Method}. In Section~\ref{Data}, we analyse the Auger data of Ref.~\cite{Collaboration:2010zzj} and report the noteworthy feature. Section~\ref{Discussion} presents a discussion of the results.

\section{Method}
\label{Method}

For UHE proton sources, events are expected to be deflected on the sky from the sources as $1/E$~\cite{Golup:2009cv,Giacinti:2009fy}. For heavy nuclei primaries, images of sources located in most regions of the sky would have more complicated shapes, even at the highest energies~\cite{Harari:1999it,Harari:2000he,Harari:2000az,Giacinti:2010dk,Giacinti:2011uj,ICRC_0229}. We find that in some GMF models and some regions of the sky, sources of nuclei of charge $Z$ can still have at least one of their images above $\sim 55$\,EeV which roughly looks like a proton source image enlarged by a factor $\sim Z$~\cite{Giacinti:2010ep}. Such images can be detected without a better knowledge of the GMF than currently available. The following method is optimized to look for such favourable heavy nuclei source images in the data.

 \begin{figure}[!t]
  \vspace{5mm}
  \centering
  \includegraphics[width=0.43\textwidth]{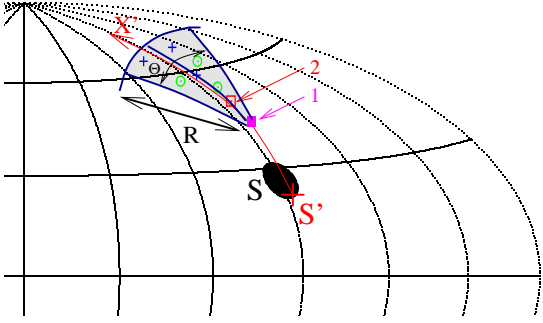}
  \caption{Sketch of the method proposed to detect extended UHE heavy nuclei sources S (black disk) with ``enlarged proton-like'' images. The source events are represented by decreasing energy order, by the magenta filled and red open boxes, green open circles and blue crosses. The sector (opening angle $\Theta$ and extension $R$) is highlighted in grey. $X'$ axis in red and S' for the reconstructed source.}
  \label{Figure1}
 \end{figure}

Fig.~\ref{Figure1} presents a sketch of the method. It starts selecting an event with energy $E_1 \geq 10^{20}$\,eV, denoted by ``1''. Then, an assumption on the UHECR deflection power $\mathcal{D}$ of the GMF~\cite{Giacinti:2009fy} is made, and only events at angular distances $\leq R = \mathcal{D}/(55\,\mbox{EeV}) - \mathcal{D}/E_{1}$ from 1 are considered hereafter. Events with energies $E_2$, located at angular distances $\leq \mathcal{D}/E_{2} - \mathcal{D}/E_{1}$ from 1, are tested by decreasing energy order. Let us take one of them, denoted ``2'' in Fig.~\ref{Figure1}. 1 and 2 define the central line of a sector shaped region, highlighted in grey in Fig.~\ref{Figure1}. The sector has an angular extension $R$ and an opening angle $\Theta$. Its vertex corresponds to 1. Let us define $X'$ as the angular distance to 1. If the number of events in the sector and their correlation coefficient Corr($X'$,$1/E$) are respectively larger or equal to $\mathcal{N}$ and $C_{\min}$, there is detection. Otherwise, the next event 2 in the ordered list is tested, until there is either detection or all events 2 have been tested without any detection. In case of detection, the source is reconstructed along the axis defined by 1 and the center of mass of all points in the sector (red axis in Fig.~\ref{Figure1}), by fitting $1/E$ versus $X'$ as in Ref.~\cite{Giacinti:2009fy}. This method has four free parameters: $\mathcal{N}$, $C_{\min}$, $\mathcal{D}$ and $\Theta$. $\mathcal{D}$ and $\Theta$ respectively depend on the regular and turbulent GMF contributions to UHECR deflections.

\section{Application to the Auger data}
\label{Data}

We apply the method presented in Section~\ref{Method} to the 69 Auger events, with $E \geq 55$\,EeV, published in Ref.~\cite{Collaboration:2010zzj}. We scan them over discretized sets of values for $\mathcal{N}$, $C_{\min}$, $\mathcal{D}$ and $\Theta$. $\mathcal{N} \in \left\lbrace 4,5,...,69\right\rbrace $, $C_{\min} \in \left\lbrace -1,-0.9,...,0.9\right\rbrace $. According to the relative contributions to UHECR deflections of the regular and turbulent GMF components found by Ref.~\cite{Tinyakov:2004pw}, $\Theta \in \left\lbrace 10^{\circ},20^{\circ},...,80^{\circ}\right\rbrace $ should suffice. Ref.~\cite{Kachelriess:2005qm} notes that deflection angles on the sky for $10^{20}$\,eV protons are expected to be $\simeq1-2.5^{\circ}$. Therefore, assuming iron nuclei, we take $\mathcal{D} \in \left\lbrace 26^{\circ}, 39^{\circ}, 52^{\circ}, 65^{\circ}\right\rbrace \times 10^{20}\,\mbox{eV}$. We confront the data with some random background, made of 69 events distributed on the sky according to the Auger exposure. Their energies are distributed within four bins: [55\,EeV,10$^{19.8}$\,eV], [10$^{19.8}$,10$^{19.9}$]\,eV, [10$^{19.9}$,10$^{20}$]\,eV and above 10$^{20}$\,eV, according to the 69 events energies. No noteworthy dependence on the distribution of energies within the bins was found. For the last bin, we take a $E^{-4.3}$ spectrum, following~\cite{Abraham:2010mj}, and a maximum energy $E_{\max}=10^{20.5}$\,eV, because the attenuation length of heavy nuclei rapidly becomes shorter than a few Mpc above this energy~\cite{Allard:2005ha}. We checked that assuming lower values for $E_{\max}$ only increases the significance of the signal detected below.

\begin{figure}[!t]
  \vspace{5mm}
  \centering
  \includegraphics[width=0.43\textwidth]{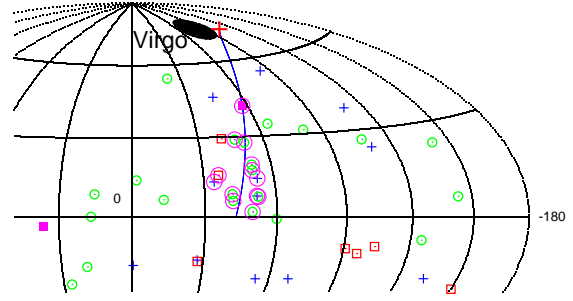}
  \caption{Portion of the sky in Galactic coordinates, showing the most significant feature detected with our method in the Auger data of Ref.~\cite{Collaboration:2010zzj}. The 13 events in the sector are surrounded with magenta circles. Symbols for the Auger cosmic rays according to their energies -see text. Black disk for the Virgo cluster, blue line for the $X'$ axis and thick red cross for the position of the reconstructed source.}
  \label{Figure2}
 \end{figure}

An interesting signal is found in the Auger data around the event with energy $E=142$\,EeV, which is located at $\simeq34^{\circ}$ from the center of Virgo. This event plays the role of ``1'' in the method presented in Fig.~\ref{Figure1}. For the tested ranges of parameters, the lowest probability to reproduce the Auger data with the background is found for $\mathcal{N}=13$, $C_{\min}=0.6$, $\mathcal{D}=39^{\circ} \times 10^{20}$\,eV and $\Theta=40^{\circ}$. In the data, Corr($X'$,$1/E)\simeq0.66$. These 13 detected events are plotted in Galactic coordinates and surrounded with magenta circles in Fig.~\ref{Figure2}. The filled magenta boxes, red open boxes, green open circles and blue crosses respectively correspond to the arrival directions on the sky of Auger events with energies $E \geq 10^{20}$\,eV, $10^{19.9}$\,eV\,$\leq E \leq 10^{20}$\,eV, $10^{19.8}$\,eV\,$\leq E \leq 10^{19.9}$\,eV and $E \leq 10^{19.8}$\,eV.

Since one cannot know \textit{a priori} the best values for the four free parameters of the method, one has to penalize over all of them~\cite{Giacinti:2010ep}. After penalization, the probability to find an at least as significant feature in the background as the feature detected in Fig.~\ref{Figure2} is $\simeq6.6\times10^{-3}$. The source associated with these 13 events (red cross in Fig.~\ref{Figure2}) is reconstructed at $\simeq 8.5^{\circ}$ from M87, near the Virgo cluster position (black disk). Even in case Virgo would only contain one UHE heavy nuclei source, it would shine as an extended source on the sky because of magnetic fields inside the cluster~\cite{virgoMF}. Moreover, due to the poor $1/E$ ordering of events expected for heavy nuclei sources, the accuracy on the position of the reconstructed source would not be better than $\pm 10^{\circ}$. Therefore, the detected events are compatible with a common emission from Virgo. (Most of) the events in the ``Cen~A region'' ($-60^{\circ}\leq l \leq -30^{\circ}$ and $0^{\circ}\leq b \leq 30^{\circ}$) and the 142\,EeV event may be the heavy nuclei image of Virgo, deflected in the GMF. If one also adds the condition that the reconstructed source should be located at less than $10^{\circ}$ from M87, the above probability falls to $\simeq3\times10^{-5}$. One may argue that this detection can have been triggered by the overdensity of events in the ``Cen~A region'' which may be due to another reason. In such a case, the relevant probability is the probability to reconstruct the source at less than $10^{\circ}$ from M87 in a data set already containing a feature as significant as the Auger feature. It is equal to $\simeq 4 \times10^{-3}$.

\section{Discussion}
\label{Discussion}

\begin{figure}[!t]
  \vspace{5mm}
  \centering
  \includegraphics[width=0.43\textwidth]{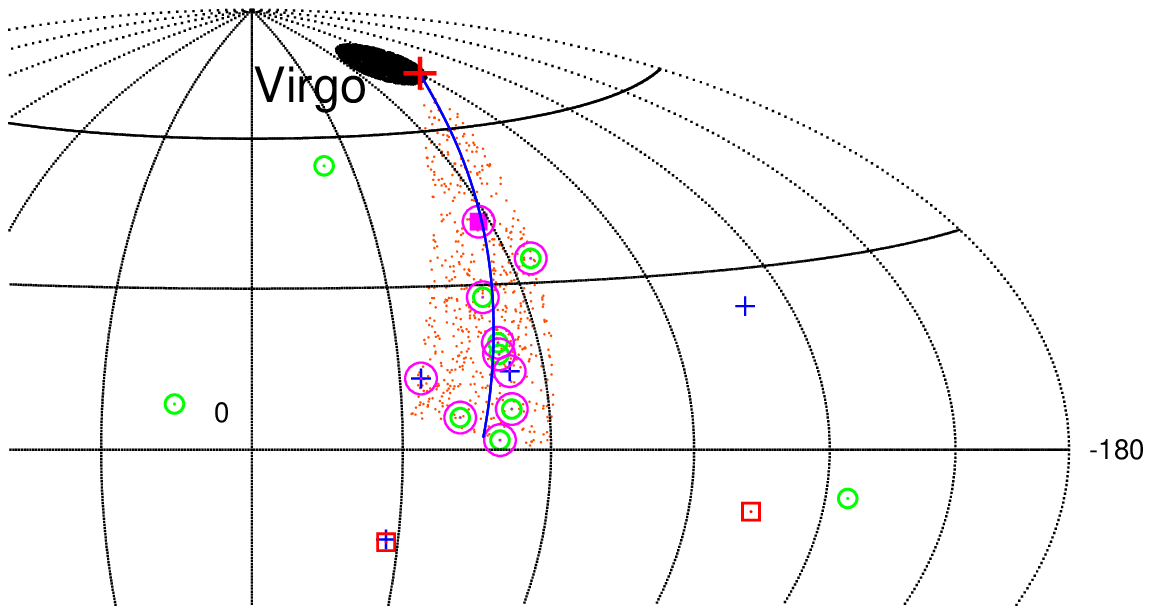}
  \includegraphics[width=0.43\textwidth]{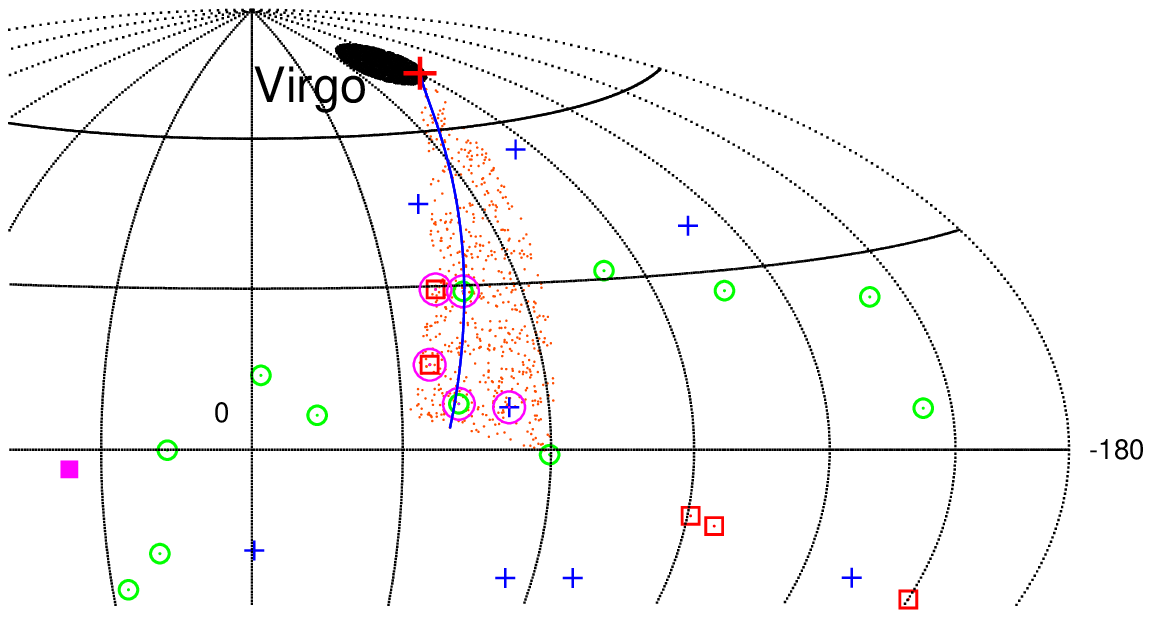}
  \caption{Blind-like analysis with the two consecutive Auger data sets (27 and $69-27=42$ events). ``Best sector'' in orange, and red cross for its vertex. Same key as in Fig.~\ref{Figure2}. \textbf{Upper panel:} First data set. The sector contains the 10 events surrounded with magenta circles; \textbf{Lower panel:} Second data set. The sector contains 5 events.}
  \label{Figure34}
 \end{figure}

We first cross-check the result of Section~\ref{Data} with a ``blind-like'' analysis by dividing the Auger data set in two: The first 27 events and the $69-27=42$ newer ones. For the first set, we determine the ``best sector'' on the sky in which the significance of the signal is the largest~\cite{Giacinti:2010ep}, compared to some random background made of 27 events following Auger exposure and spectrum. This sector is highlighted in orange in Fig.~\ref{Figure34}. It contains 10 out of the 27 events -see upper panel. Then, the newer $69-27=42$ events are analysed with this sector. It contains 5 events out of 42, and Corr($X'$,$1/E)\simeq 0.38$ -see lower panel. The probability to have in some background made of 42 events at least 5 events in this sector and Corr($X'$,$1/E) \geq 0.38$ is $\sim 2$\%. This value is compatible with the order of magnitude of the probability computed in the previous Section (0.4\%).

\begin{figure}[!t]
  \vspace{5mm}
  \centering
  \includegraphics[width=0.3\textwidth,angle=-90]{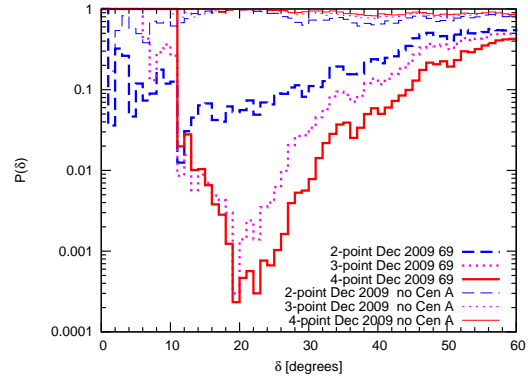}
  \caption{Probability that the signal in the 2, 3 and 4-point autocorrelation functions is a fluctuation of the background as a function of the angle $\delta$, both for the sets of 69 events (labels ``69'') and of 56 events, excluding the $20^{\circ}$ region around Cen~A (labels ``no Cen A'').}
  \label{Figure5}
 \end{figure}

We suggest in Section~\ref{Data} that the events in the Cen~A region may be a hint at the heavy nuclei image of Virgo. Let us note that the anisotropy above 55\,EeV in the Auger data is due to the overdensity of events in the Cen~A region. By computing the 2, 3 and 4-point autocorrelation functions, defined as in Ref.~\cite{Semikoz:2010cc}, we find that the rest of the sky is currently still compatible with isotropy. We plot in Fig.~\ref{Figure5} the probability that the signal in the data is a fluctuation of the background as a function of the angle (labels ``69''). The 3 and 4-point autocorrelation functions have a minimum on the $\sim 20^{\circ}$ scale. If one removes the events in the $20^{\circ}$ region around Cen~A (``no Cen A''), the minimum disappears.

\begin{figure}[!t]
  \vspace{5mm}
  \centering
  \includegraphics[width=0.43\textwidth]{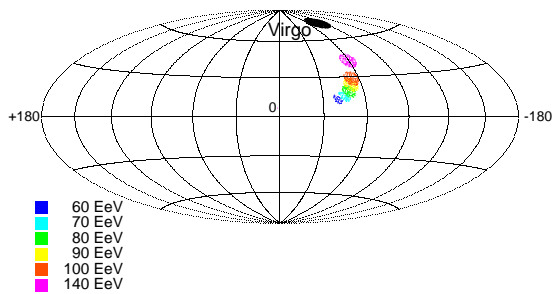}
  \caption{Image of UHE iron nuclei emitted by Virgo (black disk) and deflected in the regular GMF model discussed in~\cite{Giacinti:2010ep}. Shaded areas represent the arrival directions at Earth of cosmic rays with given energies, see key.}
  \label{Figure6}
 \end{figure}

Let us also point out that some GMF configurations are compatible with the possibility presented in Section~\ref{Data}. Fig.~\ref{Figure6} shows the iron image, for $E=60-140$\,EeV, of the Virgo cluster in a reshaped version of existing GMF models, whose parameters are discussed in~\cite{Giacinti:2010ep}. If one adds a turbulent component to this model, low energy events would be spread in the whole Cen~A region. If Virgo would be confirmed in the future to be a heavy nuclei source, it would put strong constraints on the GMF. For instance, the 142\,EeV event would give an estimate of UHECR deflections in the Galactic Northern halo, and the contribution of the turbulent GMF to deflections should be small because of the small spread of the image in the Cen~A region. It would also bring constraints on the GMF strengths, extensions and geometries in the Galactic disk and halo~\cite{Giacinti:2010ep}.

This would put restrictions on the source(s), too. Due to their equal rigidities $E/Z$, $60 - 80$\,EeV iron nuclei are deflected as $2 - 3$\,EeV protons in cosmic magnetic fields. Thus, if UHECR sources accelerate both nuclei and protons and if the Cen A region events are heavy nuclei, one should have protons at $Z$ times lower energies, in the same region~\cite{Lemoine:2009pw}. Since Auger data is compatible with isotropy at low energies, this would imply either that the source spectrum is harder than a $E^{-2}$ spectrum or that the ratio of accelerated protons to nuclei in the source(s) is not larger than one to one. The source(s) may for instance have a hard spectrum as in the model of Ref.~\cite{Neronov:2009zz}. This would also put restrictions on the maximum acceleration energy, as well as on the physical conditions in the Virgo cluster~\cite{Giacinti:2010ep}.

Only next generation experiments, such as JEM-EUSO, will have enough statistics to confirm or rule out this possibility. Auger will triple its statistics during its lifetime. If Virgo is the source of the 142\,EeV event, it may detect another event above $10^{20}$\,eV in the same region, which would be a hint. It will also settle if the Cen~A overdensity is not a statistical fluctuation. If it is not a fluctuation, and if the Auger composition is correct, there are three main possibilities: this overdensity may be due to magnetic lensing~\cite{Giacinti:2010dk,ICRC_0229}, to nuclei from Virgo as proposed above, or to nuclei emitted by Cen~A~\cite{Gorbunov:2008ef}. However, Cen~A may not be powerful enough to be the source~\cite{Casse:2001be}.

\section{Conclusions and perspectives}

We have proposed here both a new method to detect some UHE heavy nuclei sources in the data, and a new interpretation of the Auger data which is both consistent with its composition and anisotropy measurements.

In Section~\ref{Method}, we showed that some heavy nuclei sources located in some parts of the sky may have at least one of their images which looks like an enlarged proton source image. We presented a method which is able to detect such favourable sources on top of background. In Section~\ref{Data}, we applied this method to the 69 Auger events with $E>55$\,EeV of Ref.~\cite{Collaboration:2010zzj} and found that the 142\,EeV event and the Cen~A region events may be the UHE heavy nuclei image of the Virgo cluster. The associated reconstructed source is indeed located at only a few degrees from Virgo, at $\simeq 8.5^{\circ}$ from M87. The probability to reconstruct the source at less than $10^{\circ}$ from M87 in some data already containing a comparable set of events is $\sim 0.4$\%. A better knowledge of the GMF than currently available, and more UHECR data are however still needed to confirm or rule out this possibility.

In Section~\ref{Discussion}, we cross-checked this possibility with a ``blind-like'' analysis. We also pointed out that the anisotropy in the Auger data is due to the Cen~A region events. We note that the next generation of UHECR experiments will be able to test the possibility presented here.


\clearpage

\end{document}